\documentclass{article}
\usepackage[top=30mm, bottom=30mm, left=25mm, right=30mm]{geometry}
\usepackage{hyperref}
\usepackage{amsmath}
\usepackage{stmaryrd}
\usepackage{graphicx}
\usepackage{float}
\usepackage{cite}
\usepackage[affil-it]{authblk}
\begin{document}

\title{Parton distribution functions with QED corrections in the valon model}

\author{Marzieh Mottaghizadeh, Fatemeh Taghavi Shahri\thanks{Corresponding author: taghavishahri@um.ac.ir},
Parvin Eslami}

\affil{Department of Physics, Ferdowsi University of Mashhad,
Mashhad, Iran}\maketitle

‎
\begin{abstract}
Parton distribution functions (PDFs) with QED corrections extracted from the QED$\otimes$QCD DGLAP evolution equations in the framework of ``valon'' model. Our results for the PDFs with QED corrections in this phenomenological model are in good agreement with the newly related CT14QED global fit code [Phys. Rev. D93, 114015 (2016)] and APFEL (NNPDF2.3QED) [Computer Physics Communications 185, 1647 (2014)] program in a wide range of $x=[10^{-5}, 1]$ and $Q^2=[0.283, 10^8] \, {\rm GeV}^2$. The model calculations agree rather well with those codes. We also proposed the new method for studying the symmetry breaking of the sea quarks distribution functions inside proton. Then these PDFs set can be used to explore the proton-proton scattering at the LHC
era.
\end{abstract}

\section{Introduction}\label{sec:intro}

Nowadays, a deep knowledge of the properties of hadrons and precise understanding of parton distributions functions (PDFs) are key ingredients in searches for new physics at the LHC (for a review see e.g. Refs.~\cite{Brooijmans:2016vro,Bhattacherjee:2015xra}).
Hence, reliable extraction of information on the polarized PDFs~\cite{Shahri:2016uzl,Ball:2013lla,Jimenez-Delgado:2014xza,Sato:2016tuz,Nocera:2014uea,Leader:2014uua,Nocera:2014gqa,Khanpour:2017cha}, unpolarized PDFs~\cite{Ball:2017nwa,Bourrely:2015kla,Harland-Lang:2014zoa,Ball:2014uwa,Martin:2009iq,Gao:2013xoa,Khanpour:2016uxh,MoosaviNejad:2016ebo}, and nuclear PDFs~\cite{Khanpour:2016pph,Eskola:2016oht,Kovarik:2015cma,Klasen:2017kwb,Wang:2016mzo,deFlorian:2011fp} from global QCD analyses of DIS data as well as all related studies~\cite{Bertone:2017tyb,Goharipour:2017rjl,Carlson:2017gpk,Monahan:2016bvm,Dahiya:2016wjf,Jimenez-Delgado:2013boa,Nocera:2016zyg,Ball:2016spl,Goharipour:2017uic,Ru:2016wfx,Haider:2016zrk,Accardi:2016qay,Armesto:2015lrg,Frankfurt:2015cwa,Guzey:2013xba,Frankfurt:2016qca,Guzey:2016qwo,Frankfurt:2011cs,Alekhin:2017kpj,Salajegheh:2015xoa,Chen:2017mzz,Kalantarians:2017mkj,Ethier:2017zbq,Kusina:2016fxy,Boroun:2015yea,Boroun:2014nia,Boroun:2014yea,AtashbarTehrani:2013qea,TaghaviShahri:2010zz,Shoeibi:2017lrl,Kovchegov:2017lsr}, provides deep understanding on the structure of hadrons in term of their quarks and gluon constituents. With the advent of the electron-proton $(ep)$ collider HERA, the kinematic range of the DIS regime has been widely extended, allowing to achieve a much deeper understanding of the structure of nucleons.

Recently, precision achievement by ATLAS~\cite{Aad:2013iua} experiment at the Large Hadron Collider (LHC) on Drell-Yan processes shows that the size of photon-induced contribution to the dileptons invariant mass is significant. The cross-section of such process, related to the parton distribution functions of the photon, $~x\gamma(x,Q^2)$, in proton~\cite{Sadykov:2014aua,Giuli:2017tst,Giuli:2017oii,Carrazza:2013bra,Slominski:2005bw}.

It has also been shown that the precision phenomenology at the Large Hadron Collider (LHC) requires theoretical calculations which include QCD corrections and electroweak (EW) corrections~\cite{Giuli:2017oii}. An essential ingredient of these electroweak corrections is the photon parton distribution function inside proton, $~x \gamma(x,Q^2)$. 

Inconsequently, the quantum electrodynamics (QED) and electroweak (EW) corrections are important issues on many theoretical prediction at high energy at the LHC. So, at the LHC era, the determination of photon distribution function inside the proton has become important. Therefore, to imply the inclusion of QED corrections to perturbative evolution lead to additional partons in the proton. The photon distribution functions, which produced by radiation of photon from charged quarks, can be determined from the QED$\otimes$QCD DGLAP evolution equations. There are some sets of the PDFs such as the MRST20-04QED~\cite{Martin:1998sq,Martin:2004dh}, NNPDF2.3QED~\cite{Bertone:2013vaa} and CT14QED~\cite{deFlorian:2015ujt} global fits codes that incorporated the photon contribution of proton.

The goal of this analysis is to show how a simple phenomenological model, e.g. ``valon'' model, can determine the photon distribution function in the proton. The valon model was first proposed by R.C.Hwa~\cite{Hwa:1979bx,Hwa:1994uha,Hwa:2002zu} to investigate the paron distribution functions inside the proton. In this model, proton is a bound state of three "valons". Each valon is a valence quark with its associated sea quarks and glouns. The quantum number of valon is the quantum number of its valence quark and the valons carry all the momentum of proton . In this model, the recombination of parton into hadrons occur in two stage processes: at first, the partons emit and absorb glouns(and here, photon) to evolved the quark-gloun (photon) cloud and became valons. These valons then recombine into hadron.

The organization of this paper is as follows: In Section~\ref{sec2}, we bring out the QED$\otimes$QCD DGLAP evolution equations with suitable initial inputs. We also propose the novel method to study the symmetry breaking of the sea quarks distribution functions inside proton in Section~\ref{sec3}. In Section~\ref{sec:Discussion}, we discuss our findings and give conclusions on our determination of the photon PDF with QED corrections.

%
\section{QED$\otimes$QCD DGLAP evolution equations}\label{sec2}

The singlet parton distribution functions, $f_{i}(x,Q^{2})$, obey the DGLAP evolution equations~\cite{Carrazza:2015dea} in $x$ space, as

\begin{equation}
\frac{\partial}{\partial logQ^{2}}\left(\begin{array}{c}
f_{1}\\
f_{2}\\
f_{3}\\
f_{4}
\end{array}\right)=\\
\left(\begin{array}{cccc}
P_{11} & P_{12} & P_{13} & P_{14}\\
P_{21} & P_{22} & P_{23} & P_{24}\\
P_{31} & P_{32} & P_{33} & P_{34}\\
P_{41} & P_{42} & P_{43} & P_{44}
\end{array}\right)\otimes\left(\begin{array}{c}
f_{1}\\
f_{2}\\
f_{3}\\
f_{4}
\end{array}\right)\label{eq:1}
\end{equation}

and the DGLAP evolution equations for the non-singlet parton distribution functions are as follow,

\begin{equation}
\frac{\partial{{f}_{i}}}{\partial log{{Q}^{2}}}={{P}_{ii}}\otimes{{f}_{i}}\qquad i=5,\ldots,9\label{eq:2}
\end{equation}

where $P_{ij}$ and $P_{ii}$ are the splitting functions and represented in Ref.~\cite{Mottaghizadeh:2016krr} with details, and $\otimes$ denotes the convolution integral

\begin{equation}
f\otimes g=\intop_{x}^{1}\frac{dy}{y}f(y)g(\frac{x}{y})
\label{eq:3}
\end{equation}

For the coupled approach we utilize a PDF basis for the QED$\otimes$QCD DGLAP evolution equations, defined by the following singlet and non-singlet PDF combinations~\cite{Roth:2004ti},

\begin{center}
\begin{equation}
	q^{SG}:\left(\begin{array}{c}
	{{f}_{1}}=\Delta=\\
	u+\bar{u}+c+\bar{c}-d-\bar{d}-s-\bar{s}-b-\bar{b}\\
	{{f}_{2}}=\Sigma=\\
	u+\bar{u}+c+\bar{c}+d+\bar{d}+s+\bar{s}+b+\bar{b}\\
	{{f}_{3}}=g\\
	{{f}_{4}}=\gamma
	\end{array}\right)\label{eq:4}
\end{equation}
\par\end{center}

\begin{center}
\begin{equation}
	q^{NS}:\left(\begin{array}{c}
	{{f}_{5}}={{d}_{v}}=d-\bar{d}\\
	{{f}_{6}}={{u}_{v}}=u-\bar{u}\\
	{{f}_{7}}={{\Delta}_{ds}}=d+\bar{d}-s-\bar{s}\\
	{{f}_{8}}={{\Delta}_{uc}}=u+\bar{u}-c-\bar{c}\\
	{{f}_{9}}={{\Delta}_{sb}}=s+\bar{s}-b-\bar{b}
	\end{array}\right)\label{eq:5}
\end{equation}
\par\end{center}

It should be noted that the $Q^2$ dependence of the PDFs with QED corrections can be described by the QED$\otimes$QCD DGLAP evolution equations. Then, the knowledge of the PDFs at fixed scale $Q_0^2$ is enough for us to obtain the PDFs at larger scale $Q^2$. There are some solutions for the DGLAP evolutions equations with QED corrections based on the Laplace or Mellin transforms~\cite{Mottaghizadeh:2017vef,Mottaghizadeh:2016krr}. We used the solutions of these equations in Mellin space that proposed in ref. ~\cite{Mottaghizadeh:2017vef}. Here in this paper, we want to solve the QED$\otimes$QCD DGLAP evolution equations using the ``valon'' model. To solve these integro-differential evolution equations we need suitable initial inputs that simply define in this phenomenological model. So the next section is devoted to study the hadron structure in the valon model.

%
\section{Hadron structure  in the valon model}\label{sec3}

The valon model was first proposed by R.C.Hwa~\cite{Hwa:1979bx,Hwa:1994uha,Hwa:2002zu} and then
extended to the polarized structure of nucleon~\cite{Arash:2007wn,Arash:2006nw}. This model can also described the transverse structure of hadrons~\cite{Yazdi:2014zaa} and the double parton distribution functions (dPDFs)~\cite{Rinaldi:2016mlk,Gaunt:2009re} very well too.
In this model a hadron is viewed as a bound state of two or three valons. These ``constituents-quark like'' valons are defined to be the dressed valence quarks with its associated sea quarks and gluons. In scattering process, valon play a role similar to constituent quarks do  in bound state problem. It is assumed that the valons stand between partons and hadrons and the valon distributions inside hadron is universal and $Q^2$ independent. Then the proton, for example  has three valons, UUD, which carry all of the proton momentum. In the valon model, the recombination of partons into hadrons occur in two stage processes: first partons emit and
absorb gluons(photons). The evolved quarks-gluon(photon) cloud become ``valons''. Then, these valons recombine into hadrons.
Briefly, we have

\begin{itemize}
	\item At low $Q^2$, the internal structure of valons can not be resolved and the valons behave as constituent quarks.
	\item At high $Q^2$, the internal structure of valons and the $Q^2$ dependence of parton distribution functions in the hadron come from solutions of the DGLAP evolution equations in each valon with suitable initial input densities in the valon.
	\item The valon distribution functions are not depend on $Q^2$. They can be interpreted as the wave-function square of the constituent quarks in hadron. It also means the probability of finding a valon with momentum fraction of $y$ of hadron momentum.
\end{itemize}

In the following subsections, we investigate the parton distribution functions with QED corrections using the valon model.

%
\subsection{Parton distribution functions in the valon}
\begin{figure}
 \begin{minipage}{\columnwidth}
	     \centering
		\includegraphics[clip,width=0.6\textwidth]{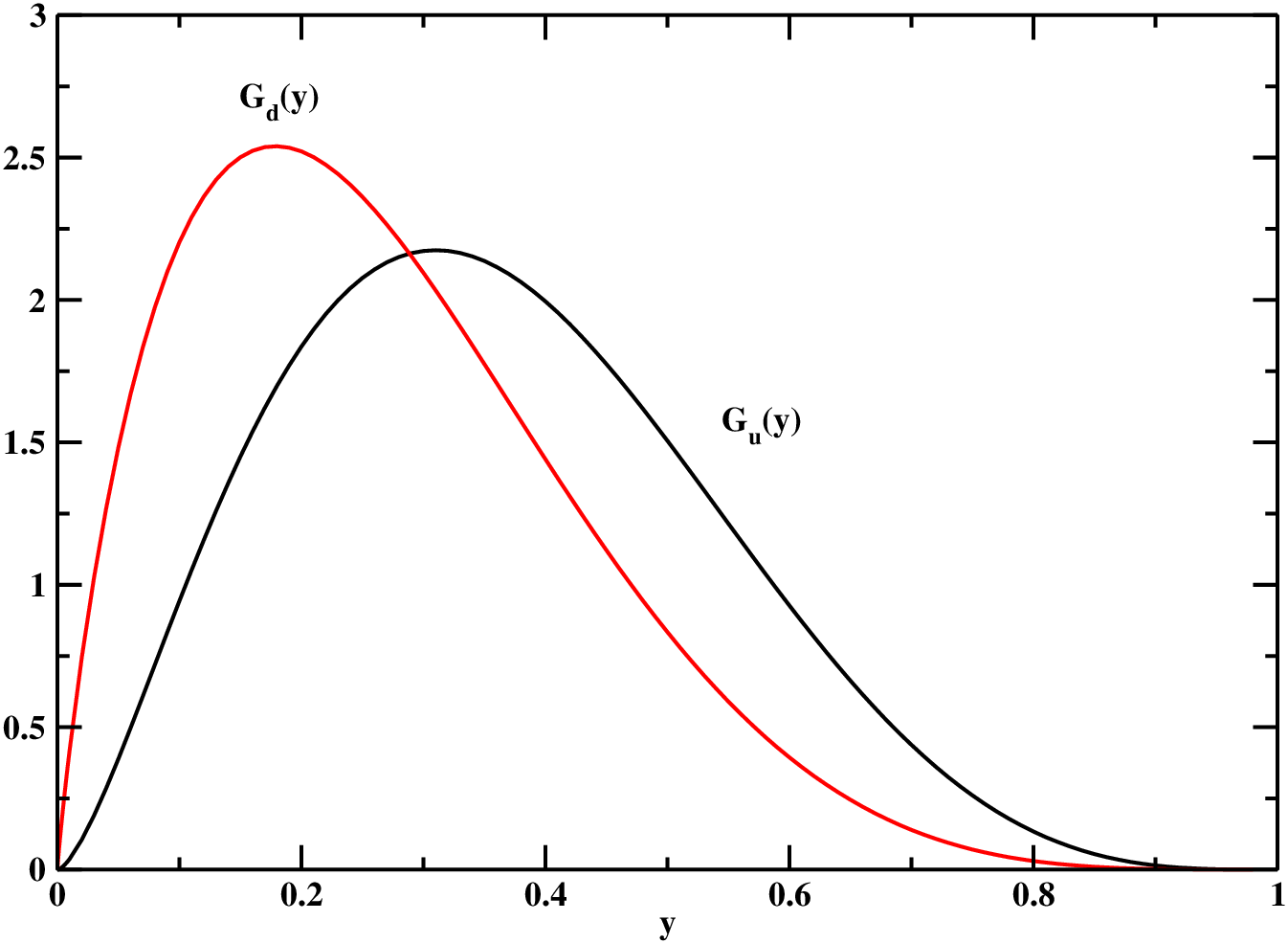}
	\end{minipage}
		\caption{{\small (color online) The valon distribution functions at $Q^2=10 \, {\rm GeV}^2$ for  U and D valons. \label{fig1}}}
\end{figure}

The valon model essentially has two steps. The first one is the solutions of the DGLAP equations with appropriate initial inputs in the valon. The second one is to convolute the parton distribution functions in the valons with the valon distribution functions, $G_{\rm valon}^{p}(y)$, to obtain the parton distributions for example in the proton, as follows

\begin{equation}
q^{p}(x,Q^{2})=\sum_{\underset{}{valon}} \intop_{x}^{1}dyG_{\rm valon}^{p}(y)q^{\rm valon}(\frac{x}{y},Q^{2})\label{eq:6}
\end{equation}
In a similar way, the proton structure function $F_{2}(x,Q^{2})$  is a convolution
of the valon distribution $G_{\rm valon}^{p}(y)$ and the structure function
of the valon, $F_{2}^{\rm valon}(z,Q^{2})$, then we have

\begin{equation}
F_{2}^{h}(x,Q^{2})=\sum_{\underset{}{valon}} \intop_{x}^{1}dyG_{\rm valon}^{h}(y)F_{2}^{valon}(\frac{x}{y},Q^{2})\label{eq:7}
\end{equation}
These valon distributions which means the probability of finding a valon with momentum fraction of $y$ of the hadron momentum are given in Refs.~\cite{Hwa:1979bx,Hwa:1994uha}. The valon distribution functions which are plotted in Fig.~\ref{fig1} are given as follows,

\begin{eqnarray}
G_{U/p}(y) & =\frac{B(\alpha+1,\beta+1)y^{\alpha}(1-y)^{\alpha+\beta+1}}{B(\alpha+1,\beta+1)B(\alpha+1,\alpha+\beta+2)}
\label{eq:8}
\end{eqnarray}

\begin{eqnarray}
G_{D/p}(y) & =\frac{B(\alpha+1,\alpha+1)y^{\beta}(1-y)^{2\alpha+1}}{B(\alpha+1,\beta+1)B(\alpha+1,\alpha+\beta+2)}
\label{eq:9}
\end{eqnarray}
where $B(m,n)$ is the beta function and $\alpha=1.545$ and $\beta=0.89$ \cite{Hwa:1979bx,Hwa:1994uha}.

\begin{figure}
    \begin{minipage}{\columnwidth}
	     \centering
		\includegraphics[clip,width=0.6\textwidth]{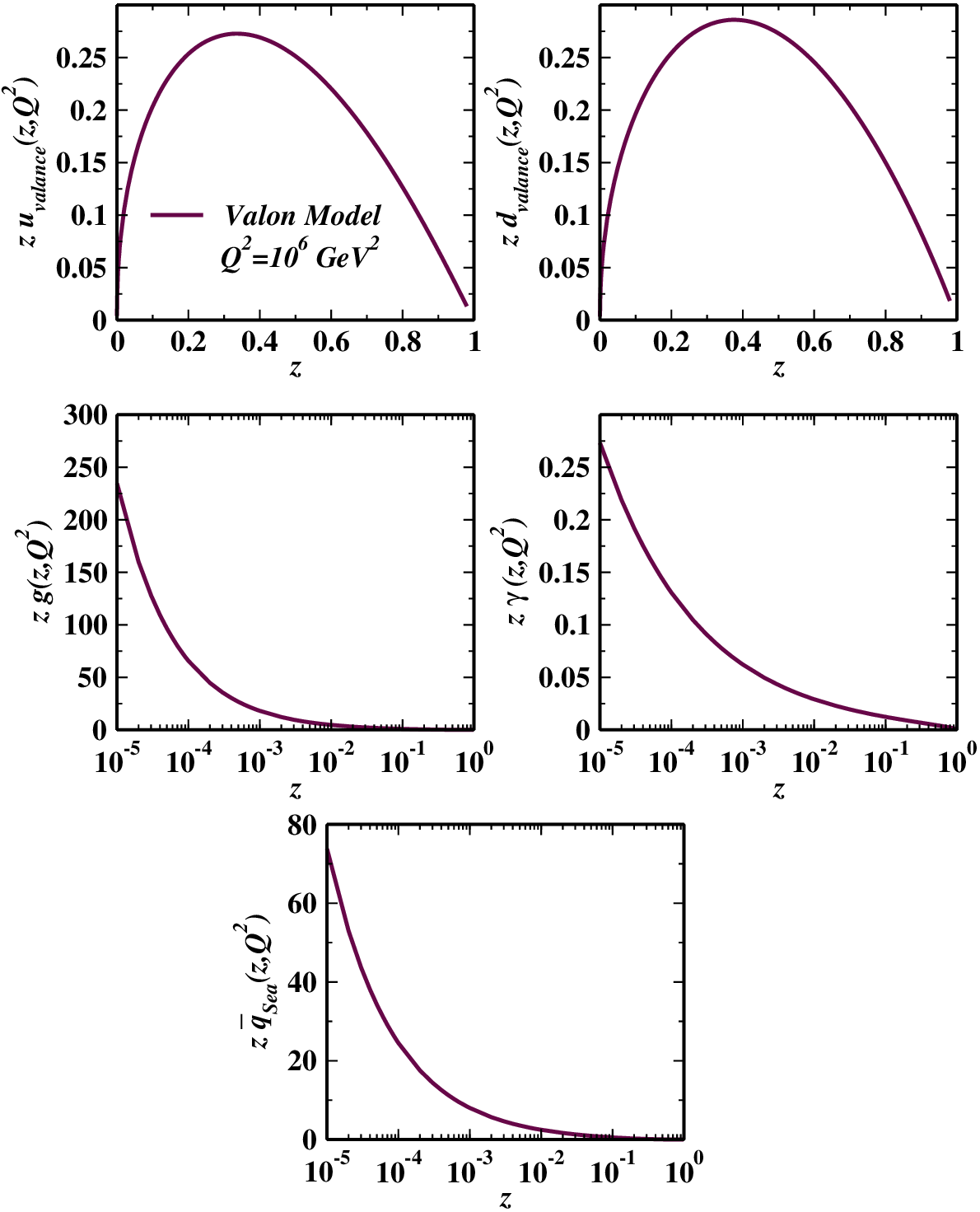}
		\end{minipage}
		\caption{{\small (color online) The parton distribution functions inside valons at $Q^{2}=10^6 \,  {\rm GeV}^{2}$ in $z$ space. \label{fig2}}}
\end{figure}

The parton distribution functions in the valon,  $q^{\rm valon}(\frac{x}{y},Q^{2})$, come from solutions of the DGLAP evolution equations in each valon. Here in this paper, we consider the parton distribution functions with QED corrections. Therefore, we solved the QED $\otimes$ QCD DGLAP evolution equations in Eqs.(~\ref{eq:1}),(~\ref{eq:2}). For to solve these equations we need initial input densities in the valons.  We work in the $\overline{MS}$ scheme with $\Lambda_{\rm QCD} = 0.22 \, {\rm GeV}$ and $Q_0^2 = 0.283 \, {\rm GeV}^2$. The
motivation for the low value of $Q_0^2$ is the phenomenological consideration to requires us to choose the initial input densities as $\delta (z-1)$  at $Q_0^2$ (where, $z=\frac{x}{y}$). This means that at such low initial scale of $Q_0^2$, the nucleon can be considered as a bound state of three valence quarks which carry all of the nucleon momentum. Therefore, at this scale of $Q_0^2$, there is one valence quark in each valon and this valence quark carry all of the valon momentum. So we should choose the initial input densities in the valon as $q^{\rm valon}(\frac{x}{y}, Q_0^2)=\delta (z-1)$ in z space. To study the evolution of the parton distribution functions, we used the solutions of the QED$\otimes$QCD DGLAP evolutions equations in Mellin space ~\cite{Mottaghizadeh:2017vef}, so that the integro-differential evolution equations reduce to sums of the parton distribution functions and pre-computable evolution kernels. Then, we choose the initial input densities for these equations in Mellin space, as follows

\begin{equation}
\begin{split}
&f_{10}=1, f_{20}=1, f_{30}=0, f_{40}=0,
\\&f_{50}=1, f_{60}=1, f_{70}=1, f_{80}=1, f_{90}=0
\label{eq:10}
\end{split}
\end{equation}
where the Mellin transform is defined as follows,
\begin{equation}
f^{N}=\intop_{0}^{1} dx x^{N-1}f(x) \label{eq:11}
\end{equation}
Our results for the parton distribution functions inside each valon are shown in Fig.~\ref{fig2}. These plots depict in $Q^2=10^6 \,{\rm GeV}^2$ as a
function of $z$.

\begin{figure}
	 \begin{minipage}{\columnwidth}
	     \centering
		\includegraphics[clip,width=0.6\textwidth]{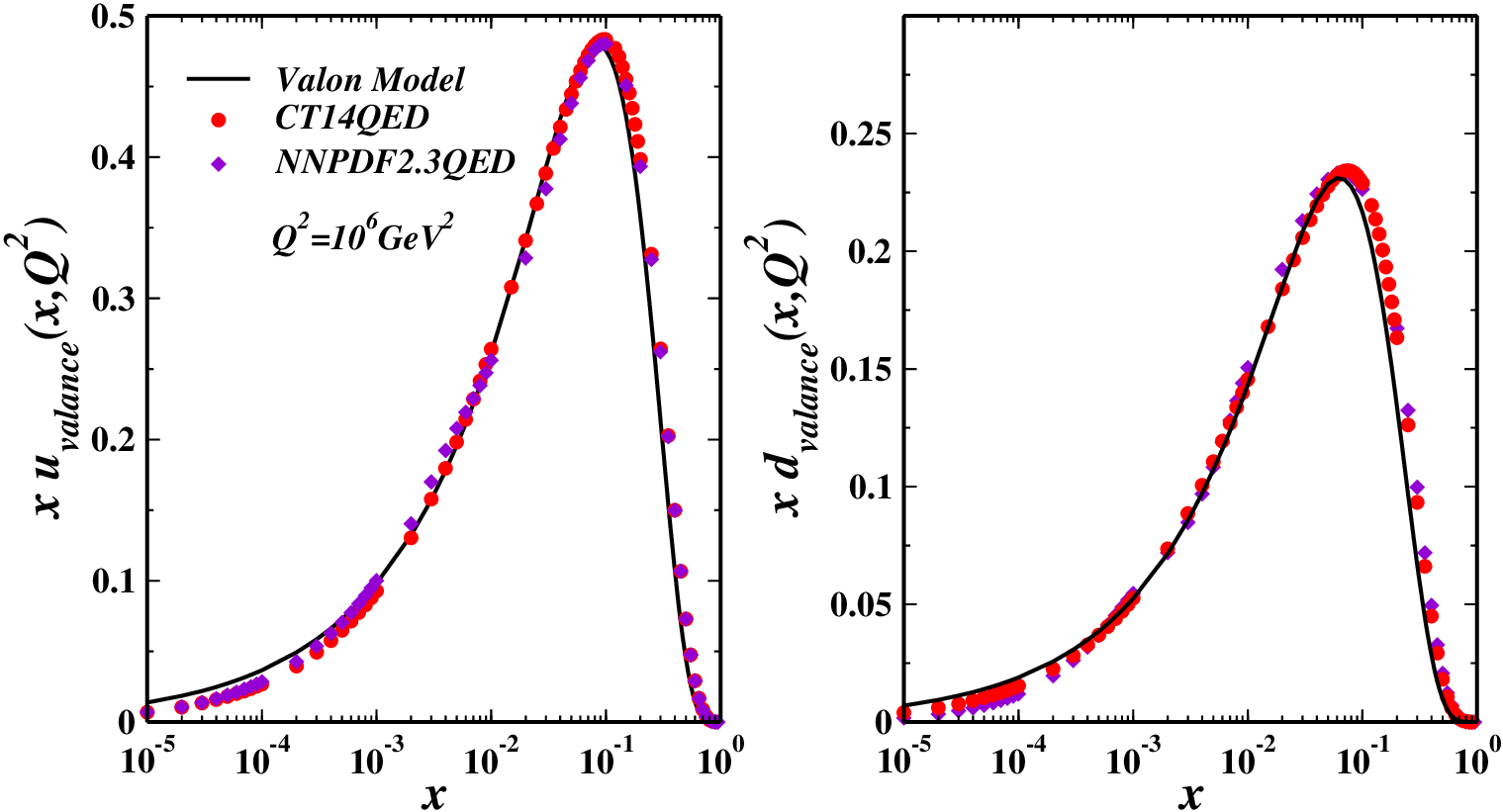}
		\end{minipage}
		\caption{{\small (color online) The parton distribution $xu_{valance}(x,Q^{2})$ and $xd_{valance}(x,Q^{2})$
				at $Q^{2}=10^6 GeV^{2}$. The solid line is our results in the
				valon model, diamond scatter is the APFEL (NNPDF2.3QED) program ~\cite{Bertone:2013vaa},
				circle scatter is the CT14QED code~\cite{deFlorian:2015ujt}. \label{fig3}}}
\end{figure}
\begin{figure*}[!h]
\centering
    	\includegraphics[clip,width=15 cm]{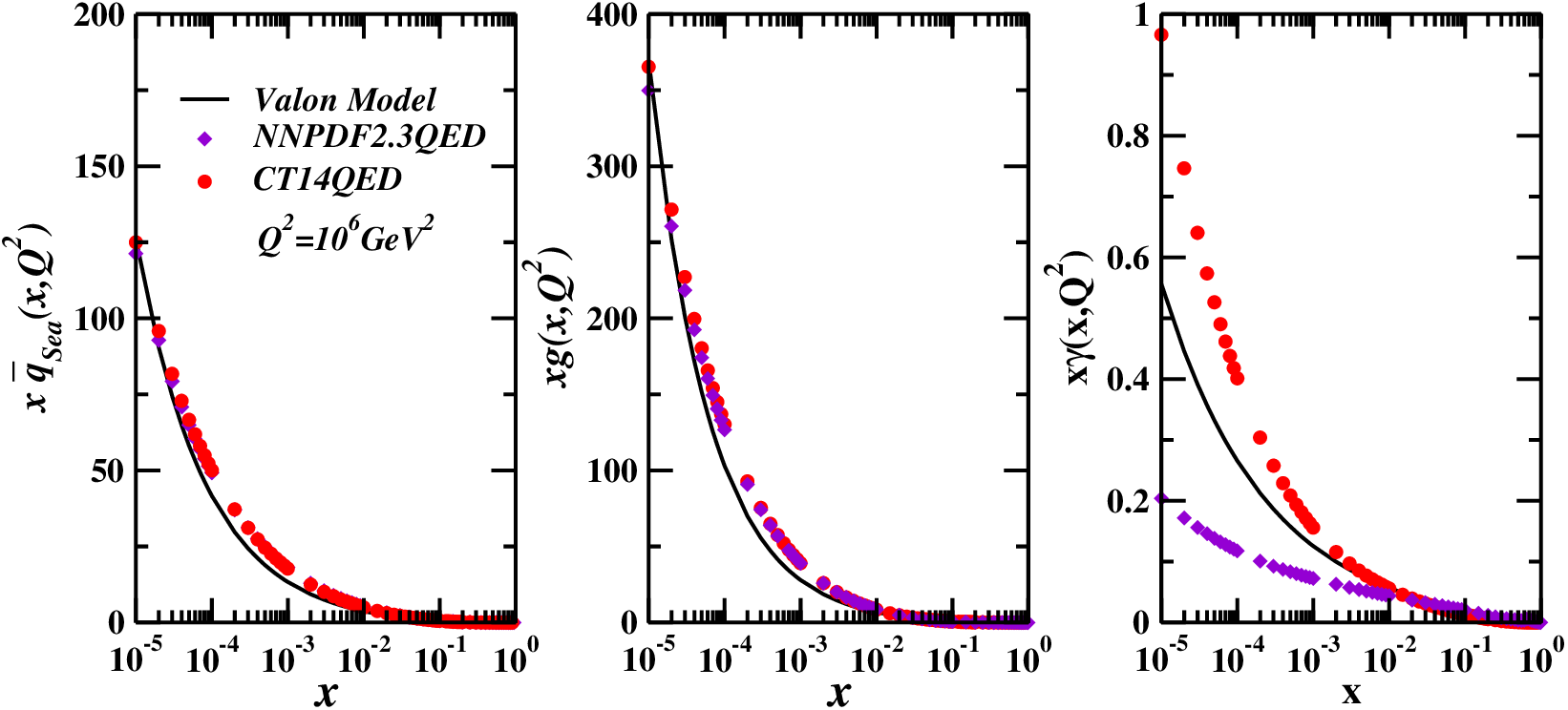}

		\caption{{\small (color online) The total sea quarks, $\bar{q}_{\rm Sea}(x, Q^2)$, gloun and photon distribution functions at $Q^2=10^6 \, {\rm GeV}^2$.
				The solid lines are our results in the valon model; diamond scatter is the APFEL (NNPDF2.3QED) code~\cite{Bertone:2013vaa}, circle scatter is the
				CT14QED code ~\cite{deFlorian:2015ujt}. \label{fig4}}}
\end{figure*}

Finally, the convolution integral of Eq.(~\ref{eq:6}) led us to the parton distribution functions with QED corrections inside  proton. The valance quark distribution functions is shown in Fig. ~\ref{fig3} at $Q^2=10^6 \, {\rm GeV}^2$ in valon model. In this figure, we compare our results with the PDFs extracted from the CT14QED global fits code and APFEL (NNPDF2.3QED) program. An excellent agreement is found for all of the flavours. Also, it is found that with an increasing in the values of $Q^2$, the valance quark distribution functions decrease for all of the values of x. Then, the contribution of photon distribution function increase with an increasing in $Q^2$.
The total sea quarks , $\bar{q}_{\rm Sea}(x, Q^2)$, gluon and photon distribution functions are plotted in Fig.~\ref{fig4}. Here we have good agreement with those from  the CT14QED global fits code and APFEL (NNPDF2.3QED) program too.


%
\subsection{Symmetry breaking in the sea quarks distribution functions}

In this subsections, we would like to know how can separate the different kind of sea quarks distributions when we know the total distribution of all sea quarks. Here, we propose the new method based on the sea quarks mass ratio. The total sea quarks distribution functions is obtained as follows,

\begin{equation}
\begin{split}
\bar{q}_{Sea}(x,Q^{2})=&2\bar{u}(x,Q^{2})+2\bar{d}(x,Q^{2})+2\bar{s}(x,Q^{2})+2\bar{c}(x,Q^{2})\\
&+2\bar{b}(x,Q^{2})
\label{eq:12}
\end{split}
\end{equation}
where, we consider $s=\bar{s}$, $c=\bar{c}$ and $b=\bar{b}$. For to study of the symmetry breaking of sea quarks distribution functions, we use the fact that probability of finding heavier partons inside proton is smaller than those of light partons, it means:

\begin{eqnarray}
\frac{q_{i}(x,Q^{2})}{q_{j}(x,Q^{2})} & \simeq\frac{m_{j}}{m_{i}}\label{eq:13}
\end{eqnarray}
Therefore, we can calculated the bottom quark distribution function, as an example, as follows

\begin{equation}
\begin{split}
\bar{q}_{Sea}(x,Q^{2})=&2\frac{m_{b}}{m_{u}}\bar{b}(x,Q^{2})+2\frac{m_{b}}{m_{d}}\bar{b}(x,Q^{2})+2\frac{m_{b}}{m_{s}}\bar{b}(x,Q^{2})\\
&+2\frac{m_{b}}{m_{c}}\bar{b}(x,Q^{2})+2\bar{b}(x,Q^{2})
\label{eq:14}
\end{split}
\end{equation}
Then, we have

\begin{eqnarray}
\bar{b}(x,Q^{2}) & =\frac{\bar{q}_{Sea}(x,Q^{2})}{2m_{b}(\frac{1}{m_{u}}+\frac{1}{m_{d}}+\frac{1}{m_{s}}+\frac{1}{m_{c}}+\frac{1}{m_{b}})}
\label{eq:15}
\end{eqnarray}
This leads to the following general relation:

\begin{eqnarray}
\bar{q}(x,Q^{2}) & =A\frac{\bar{q}_{Sea}(x,Q^{2})}{B}
\label{eq:16}
\end{eqnarray}
The B parameter is constant for each kind of sea quarks:

\begin{equation}
B_{j}=2m_{j}(\frac{1}{m_{u}}+\frac{1}{m_{d}}+\frac{1}{m_{s}}+\frac{1}{m_{c}}+\frac{1}{m_{b}})
\label{eq:17}
\end{equation}
where j index run over all of the sea quarks flavors.

The free parameter A can be extracted from experimental data. Here, we used the CT14QED PDFs set to determine these parameters for various kind of the sea quarks. The values of A and B parameters are given in Table~\ref{table1}.

\begin{table}
	\caption{The parameters A and B for the sea quarks distribution functions.}
	\begin{centering}
		\begin{tabular}{c|c|c}
			\hline
			quarks & A & B\tabularnewline
			\hline
			\hline
			$\bar{u}$ & 4.08513 & 0.5\tabularnewline
			\hline
			$\bar{d}$ & 4.08513 & 0.5\tabularnewline
			\hline
			$\bar{s}$ & 106.213 & 13\tabularnewline
			\hline
			$\bar{c}$ & 1297.03 & 150\tabularnewline
			\hline
			$\bar{b}$ & 4289.38 & 450\tabularnewline
			\hline
		\end{tabular}
		\par\end{centering}
	\label{table1}
\end{table}

\begin{figure}
 \begin{minipage}{\columnwidth}
     \centering
		\includegraphics[clip,width=0.6\textwidth]{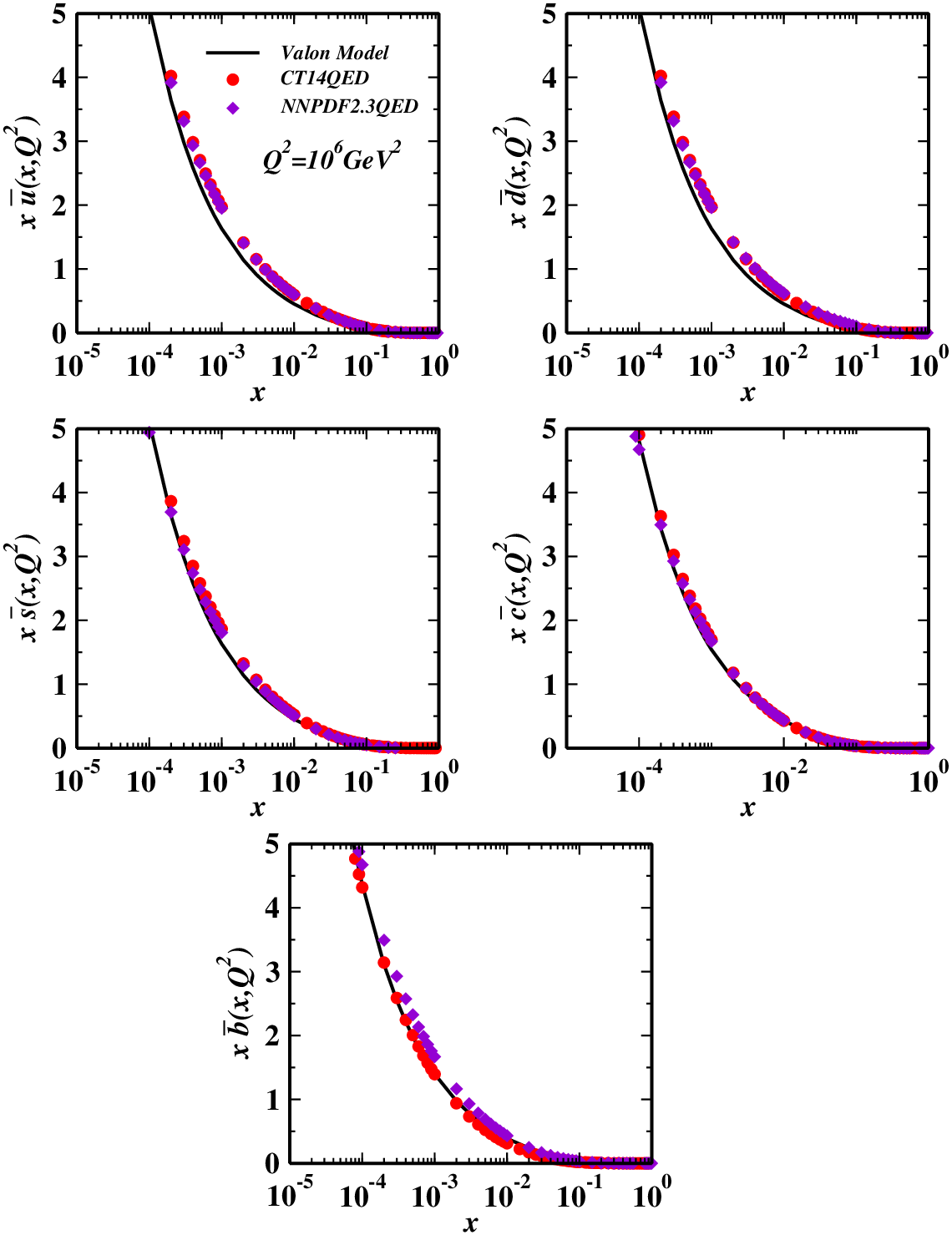}
	 \end{minipage}	
		\caption{{\small (color online) The sea quark distribution functions at $Q^{2}=10^6 \, {\rm GeV}^2$. The
				solid lines are our results from the valon model; diamond scatter 
				is the APFEL (NNPDF2.3QED) code~\cite{Bertone:2013vaa}, circle scatter is the CT14QED code~\cite{deFlorian:2015ujt}. \label{fig5}}}

\end{figure}

The sea quarks distribution functions are shown in Fig.~\ref{fig5}, they are compared with the CT14QED and APFEL (NNPDF-2.3QED) PDF sets. This figure shows that a good agreement is achieved. 

\begin{figure} 
     \begin{minipage}{\columnwidth}
     \centering
           	\includegraphics[clip,width=0.6\textwidth]{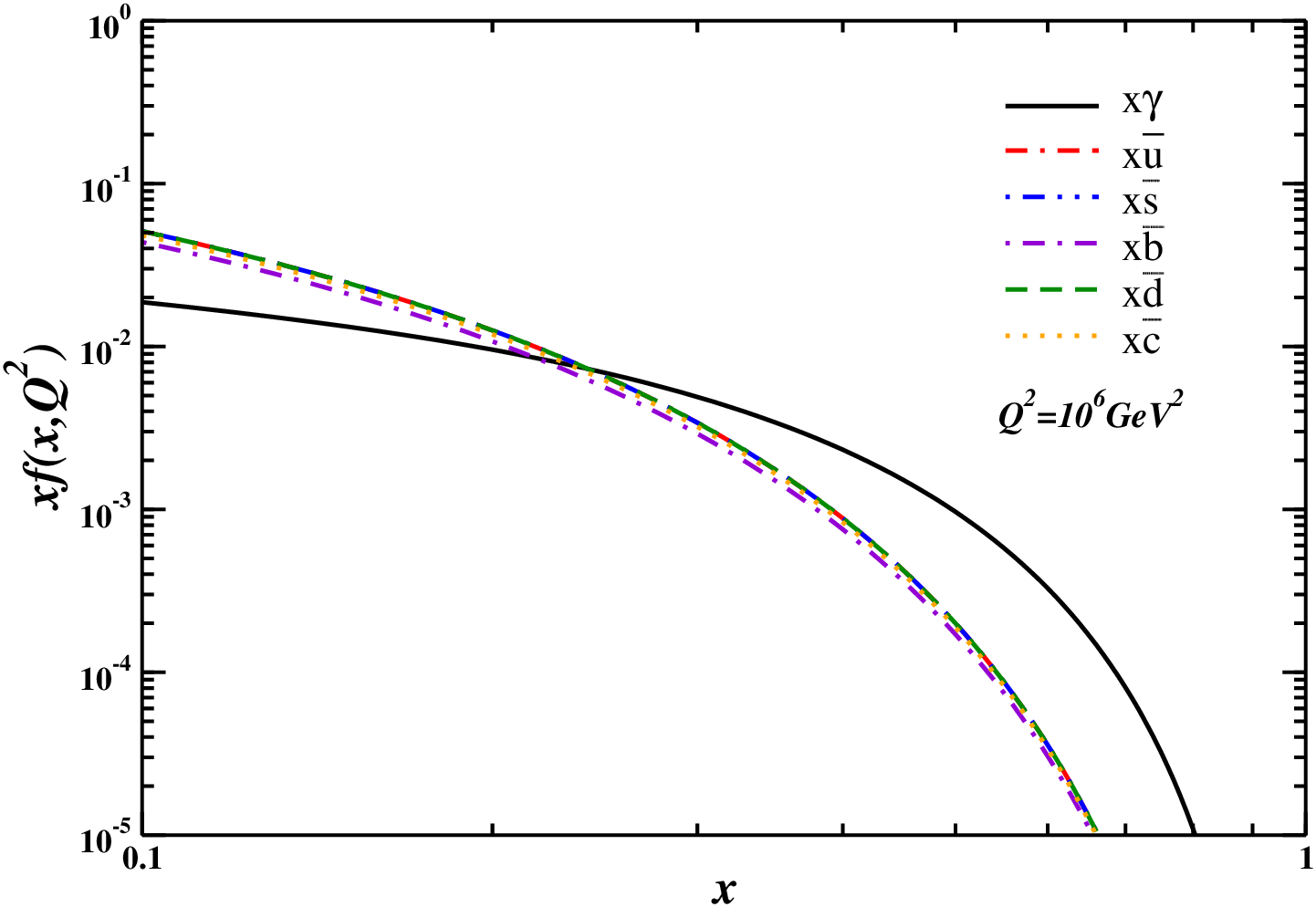}
     \end{minipage}
      \caption{The Sea quarks and photon distribution functions at $Q^{2}=10^6 \, {\rm GeV}^2$ as a function of $x$.}
\label{fig6}
\end{figure}

\begin{figure}[H]
    \begin{minipage}{\columnwidth}
    \centering
		\includegraphics[clip,width=0.6\textwidth]{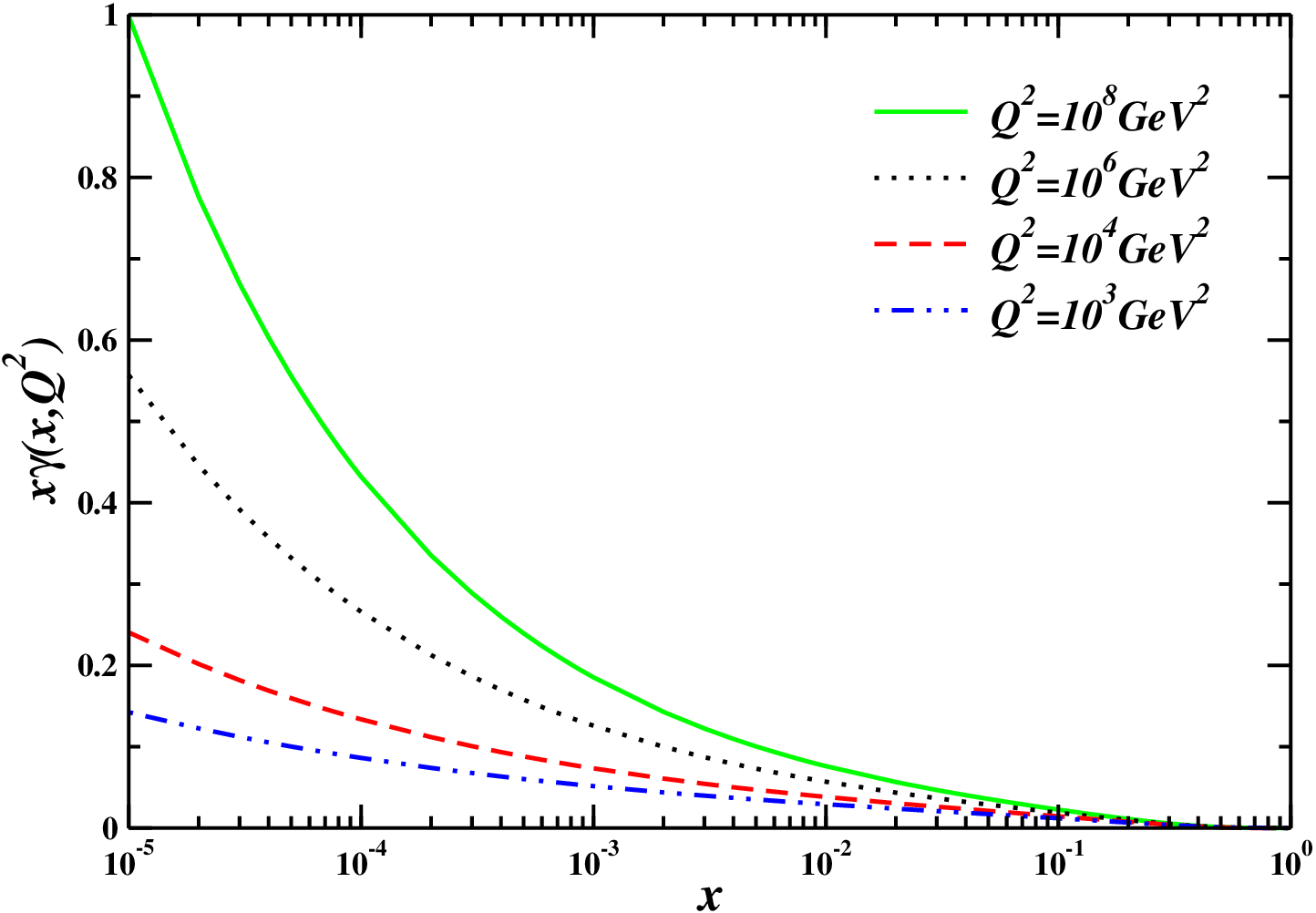}
	\end{minipage}
		\caption{{\small (color online) The photon distribution functions at different values of $Q^2$ as a function of $x$. \label{fig7}}}

\end{figure}

Fig.~\ref{fig6} shows the sea quarks and photon distribution functions in x space at fixed scale of $Q^{2}=10^{6} \, {\rm GeV}^{2}$. It is worth to notice that, the photon distribution functions are larger than the sea quarks distribution functions at large scale of energy for the large values of $x$. The photon distribution functions at different values of $Q^2$ are plotted in Fig.~\ref{fig7}. It is obvious that the photon distribution functions become more significant at high $Q^2$ where more photon are produced through radiation of the quarks.

%
\section{Summary and conclusions} \label{sec:Discussion}
%

In this paper, we calculated the PDFs with QED corrections in the valon model. The QED corrections to the parton distribution functions are important especially at high $Q^2$ where the photons can produce more partons. The importance of these PDFs set is to study the proton-proton scattering at TeV scale of energy. In the valon model, the valance quarks inside proton can emit and absorb glouns and photons and became valons. These valons can recombine into hadrons. The valon distribution functions are universal and $Q^2$ dependent. The $Q^2$ dependence of the parton distribution functions come from the solutions of the QED$\otimes$QCD DGLAP evolution equations in each valon with suitable initial inputs. We compared our QED$\otimes$QCD PDFs set
with those of the CT14QED global fits code and APFEL (NNPDF2.3QED) program. There is a nice agreement between them. The results show that the photon distribution functions are larger than the sea quarks distribution functions at high $Q^2$ and high values of $x$. The result emphasis that this simple phenomenological model can predict the hadron structures very well. The higher order QCD and QED corrections can now be added to the QED$\otimes$QCD DGLAP evolution equations inside valons. Therefore, we could extract the QED$\otimes$QCD PDFs at ${\rm N^2LO}$, ${\rm N^3LO}$,... approximations in QCD and NLO,... approximations in QED too.

%
\section*{Acknowledgments}

The authors would like to thank H. Khanpour for carefully reading the manuscript, fruitful discussion and critical remarks.

%

\begin{thebibliography}{}
%





\bibitem{Brooijmans:2016vro} 
G.~Brooijmans {\it et al.},
arXiv:1605.02684 [hep-ph].





\bibitem{Bhattacherjee:2015xra} 
B.~Bhattacherjee, T.~Modak, S.~K.~Patra and R.~Sinha,
arXiv:1503.08924 [hep-ph].





\bibitem{Shahri:2016uzl} 
F.~Taghavi-Shahri, H.~Khanpour, S.~Atashbar Tehrani and Z.~Alizadeh Yazdi,
Phys.\ Rev.\ D {\bf 93}, no. 11, 114024 (2016)
doi:10.1103/PhysRevD.93.114024
[arXiv:1603.03157 [hep-ph]].






\bibitem{Ball:2013lla} 
R.~D.~Ball {\it et al.} [NNPDF Collaboration],
Nucl.\ Phys.\ B {\bf 874}, 36 (2013)
doi:10.1016/j.nuclphysb.2013.05.007
[arXiv:1303.7236 [hep-ph]].




\bibitem{Jimenez-Delgado:2014xza} 
P.~Jimenez-Delgado {\it et al.} [Jefferson Lab Angular Momentum (JAM) Collaboration],
Phys.\ Lett.\ B {\bf 738}, 263 (2014)
doi:10.1016/j.physletb.2014.09.049
[arXiv:1403.3355 [hep-ph]].




\bibitem{Sato:2016tuz} 
N.~Sato {\it et al.} [Jefferson Lab Angular Momentum Collaboration],
Phys.\ Rev.\ D {\bf 93}, no. 7, 074005 (2016)
doi:10.1103/PhysRevD.93.074005
[arXiv:1601.07782 [hep-ph]].



\bibitem{Nocera:2014uea} 
E.~R.~Nocera,
Phys.\ Lett.\ B {\bf 742}, 117 (2015)
doi:10.1016/j.physletb.2015.01.021
[arXiv:1410.7290 [hep-ph]].




\bibitem{Leader:2014uua} 
E.~Leader, A.~V.~Sidorov and D.~B.~Stamenov,
Phys.\ Rev.\ D {\bf 91}, no. 5, 054017 (2015)
doi:10.1103/PhysRevD.91.054017
[arXiv:1410.1657 [hep-ph]].




\bibitem{Nocera:2014gqa} 
E.~R.~Nocera {\it et al.} [NNPDF Collaboration],
Nucl.\ Phys.\ B {\bf 887}, 276 (2014)
doi:10.1016/j.nuclphysb.2014.08.008
[arXiv:1406.5539 [hep-ph]].




\bibitem{Khanpour:2017cha} 
H.~Khanpour, S.~T.~Monfared and S.~Atashbar Tehrani,
Phys.\ Rev.\ D {\bf 95}, no. 7, 074006 (2017)
doi:10.1103/PhysRevD.95.074006
[arXiv:1703.09209 [hep-ph]].





\bibitem{Ball:2017nwa} 
R.~D.~Ball {\it et al.} [NNPDF Collaboration],
arXiv:1706.00428 [hep-ph].




\bibitem{Bourrely:2015kla} 
C.~Bourrely and J.~Soffer,
Nucl.\ Phys.\ A {\bf 941}, 307 (2015)
doi:10.1016/j.nuclphysa.2015.06.018
[arXiv:1502.02517 [hep-ph]].




\bibitem{Harland-Lang:2014zoa} 
L.~A.~Harland-Lang, A.~D.~Martin, P.~Motylinski and R.~S.~Thorne,
Eur.\ Phys.\ J.\ C {\bf 75}, no. 5, 204 (2015)
doi:10.1140/epjc/s10052-015-3397-6
[arXiv:1412.3989 [hep-ph]].






\bibitem{Ball:2014uwa} 
R.~D.~Ball {\it et al.} [NNPDF Collaboration],
JHEP {\bf 1504}, 040 (2015)
doi:10.1007/JHEP04(2015)040
[arXiv:1410.8849 [hep-ph]].





\bibitem{Martin:2009iq} 
A.~D.~Martin, W.~J.~Stirling, R.~S.~Thorne and G.~Watt,
Eur.\ Phys.\ J.\ C {\bf 63}, 189 (2009)
doi:10.1140/epjc/s10052-009-1072-5
[arXiv:0901.0002 [hep-ph]].





\bibitem{Gao:2013xoa} 
J.~Gao {\it et al.},
Phys.\ Rev.\ D {\bf 89}, no. 3, 033009 (2014)
doi:10.1103/PhysRevD.89.033009
[arXiv:1302.6246 [hep-ph]].




\bibitem{Khanpour:2016uxh} 
H.~Khanpour, A.~Mirjalili and S.~Atashbar Tehrani,
Phys.\ Rev.\ C {\bf 95}, no. 3, 035201 (2017)
doi:10.1103/PhysRevC.95.035201
[arXiv:1601.03508 [hep-ph]].


\bibitem{MoosaviNejad:2016ebo} 
S.~M.~Moosavi Nejad, H.~Khanpour, S.~Atashbar Tehrani and M.~Mahdavi,
Phys.\ Rev.\ C {\bf 94}, no. 4, 045201 (2016)
doi:10.1103/PhysRevC.94.045201
[arXiv:1609.05310 [hep-ph]].




\bibitem{Khanpour:2016pph} 
H.~Khanpour and S.~Atashbar Tehrani,
Phys.\ Rev.\ D {\bf 93}, no. 1, 014026 (2016)
doi:10.1103/PhysRevD.93.014026
[arXiv:1601.00939 [hep-ph]].





\bibitem{Eskola:2016oht} 
K.~J.~Eskola, P.~Paakkinen, H.~Paukkunen and C.~A.~Salgado,
arXiv:1612.05741 [hep-ph].



\bibitem{Kovarik:2015cma} 
K.~Kovarik {\it et al.},
Phys.\ Rev.\ D {\bf 93}, no. 8, 085037 (2016)
doi:10.1103/PhysRevD.93.085037
[arXiv:1509.00792 [hep-ph]].



\bibitem{Klasen:2017kwb} 
M.~Klasen, K.~Kovarik and J.~Potthoff,
arXiv:1703.02864 [hep-ph].



\bibitem{Wang:2016mzo} 
R.~Wang, X.~Chen and Q.~Fu,
arXiv:1611.03670 [hep-ph].





\bibitem{deFlorian:2011fp} 
D.~de Florian, R.~Sassot, P.~Zurita and M.~Stratmann,
Phys.\ Rev.\ D {\bf 85}, 074028 (2012)
doi:10.1103/PhysRevD.85.074028
[arXiv:1112.6324 [hep-ph]].




\bibitem{Bertone:2017tyb} 
V.~Bertone, S.~Carrazza, N.~P.~Hartland, E.~R.~Nocera and J.~Rojo,
arXiv:1706.07049 [hep-ph].





\bibitem{Goharipour:2017rjl} 
M.~Goharipour and H.~Mehraban,
arXiv:1703.01682 [hep-ph].




\bibitem{Carlson:2017gpk} 
C.~E.~Carlson and M.~Freid,
arXiv:1702.05775 [hep-ph].




\bibitem{Monahan:2016bvm} 
C.~Monahan and K.~Orginos,
arXiv:1612.01584 [hep-lat].






\bibitem{Dahiya:2016wjf} 
H.~Dahiya and M.~Randhawa,
Phys.\ Rev.\ D {\bf 93}, no. 11, 114030 (2016)
doi:10.1103/PhysRevD.93.114030
[arXiv:1606.06441 [hep-ph]].



\bibitem{Jimenez-Delgado:2013boa} 
P.~Jimenez-Delgado, A.~Accardi and W.~Melnitchouk,
Phys.\ Rev.\ D {\bf 89}, no. 3, 034025 (2014)
doi:10.1103/PhysRevD.89.034025
[arXiv:1310.3734 [hep-ph]].






\bibitem{Nocera:2016zyg} 
E.~R.~Nocera and E.~Santopinto,
arXiv:1611.07980 [hep-ph].





\bibitem{Ball:2016spl} 
R.~D.~Ball, E.~R.~Nocera and J.~Rojo,
Eur.\ Phys.\ J.\ C {\bf 76}, no. 7, 383 (2016)
doi:10.1140/epjc/s10052-016-4240-4
[arXiv:1604.00024 [hep-ph]].




\bibitem{Goharipour:2017uic} 
M.~Goharipour and H.~Mehraban,
Phys.\ Rev.\ D {\bf 95}, no. 5, 054002 (2017)
doi:10.1103/PhysRevD.95.054002
[arXiv:1702.05738 [hep-ph]].





\bibitem{Ru:2016wfx} 
P.~Ru, S.~A.~Kulagin, R.~Petti and B.~W.~Zhang,
Phys.\ Rev.\ D {\bf 94}, no. 11, 113013 (2016)
doi:10.1103/PhysRevD.94.113013
[arXiv:1608.06835 [nucl-th]].




\bibitem{Haider:2016zrk} 
H.~Haider, F.~Zaidi, M.~Sajjad Athar, S.~K.~Singh and I.~Ruiz Simo,
Nucl.\ Phys.\ A {\bf 955}, 58 (2016)
doi:10.1016/j.nuclphysa.2016.06.006
[arXiv:1603.00164 [nucl-th]].



\bibitem{Accardi:2016qay} 
A.~Accardi, L.~T.~Brady, W.~Melnitchouk, J.~F.~Owens and N.~Sato,
Phys.\ Rev.\ D {\bf 93}, no. 11, 114017 (2016)
doi:10.1103/PhysRevD.93.114017
[arXiv:1602.03154 [hep-ph]].




\bibitem{Armesto:2015lrg} 
N.~Armesto, H.~Paukkunen, J.~M.~Penï¿œn, C.~A.~Salgado and P.~Zurita,
Eur.\ Phys.\ J.\ C {\bf 76}, no. 4, 218 (2016)
doi:10.1140/epjc/s10052-016-4078-9
[arXiv:1512.01528 [hep-ph]].





\bibitem{Frankfurt:2015cwa} 
L.~Frankfurt, V.~Guzey, M.~Strikman and M.~Zhalov,
Phys.\ Lett.\ B {\bf 752}, 51 (2016)
doi:10.1016/j.physletb.2015.11.012
[arXiv:1506.07150 [hep-ph]].




\bibitem{Guzey:2013xba} 
V.~Guzey, E.~Kryshen, M.~Strikman and M.~Zhalov,
Phys.\ Lett.\ B {\bf 726}, 290 (2013)
doi:10.1016/j.physletb.2013.08.043
[arXiv:1305.1724 [hep-ph]].




\bibitem{Frankfurt:2016qca} 
L.~Frankfurt, V.~Guzey and M.~Strikman,
arXiv:1612.08273 [hep-ph].




\bibitem{Guzey:2016qwo} 
V.~Guzey, M.~Strikman and M.~Zhalov,
Phys.\ Rev.\ C {\bf 95}, no. 2, 025204 (2017)
doi:10.1103/PhysRevC.95.025204
[arXiv:1611.05471 [hep-ph]].





\bibitem{Frankfurt:2011cs} 
L.~Frankfurt, V.~Guzey and M.~Strikman,
Phys.\ Rept.\  {\bf 512}, 255 (2012)
doi:10.1016/j.physrep.2011.12.002
[arXiv:1106.2091 [hep-ph]].




\bibitem{Alekhin:2017kpj} 
S.~Alekhin, J.~Blï¿œmlein, S.~Moch and R.~Placakyte,
arXiv:1701.05838 [hep-ph].



\bibitem{Salajegheh:2015xoa} 
M.~Salajegheh,
Phys.\ Rev.\ D {\bf 92}, no. 7, 074033 (2015)
doi:10.1103/PhysRevD.92.074033
[arXiv:1602.00154 [hep-ph]].



\bibitem{Chen:2017mzz} 
J.~W.~Chen, T.~Ishikawa, L.~Jin, H.~W.~Lin, Y.~B.~Yang, J.~H.~Zhang and Y.~Zhao,
arXiv:1706.01295 [hep-lat].



\bibitem{Kalantarians:2017mkj} 
N.~Kalantarians, E.~Christy and C.~Keppel,
arXiv:1706.02002 [hep-ph].



\bibitem{Ethier:2017zbq} 
J.~J.~Ethier, N.~Sato and W.~Melnitchouk,
arXiv:1705.05889 [hep-ph].



\bibitem{Kusina:2016fxy} 
A.~Kusina {\it et al.},
arXiv:1610.02925 [nucl-th].




\bibitem{Boroun:2015yea} 
G.~R.~Boroun,
Phys.\ Lett.\ B {\bf 744}, 142 (2015)
doi:10.1016/j.physletb.2015.03.051
[arXiv:1503.01590 [hep-ph]].



\bibitem{Boroun:2014nia} 
G.~R.~Boroun,
Nucl.\ Phys.\ B {\bf 884}, 684 (2014)
doi:10.1016/j.nuclphysb.2014.05.010
[arXiv:1406.0061 [hep-ph]].



\bibitem{Boroun:2014yea} 
G.~R.~Boroun,
Phys.\ Lett.\ B {\bf 741}, 197 (2015)
doi:10.1016/j.physletb.2014.12.039
[arXiv:1411.6492 [hep-ph]].



\bibitem{AtashbarTehrani:2013qea} 
S.~Atashbar Tehrani, F.~Taghavi-Shahri, A.~Mirjalili and M.~M.~Yazdanpanah,
Phys.\ Rev.\ D {\bf 87}, no. 11, 114012 (2013)
Erratum: [Phys.\ Rev.\ D {\bf 88}, no. 3, 039902 (2013)].
doi:10.1103/PhysRevD.87.114012, 10.1103/PhysRevD.88.039902



\bibitem{TaghaviShahri:2010zz} 
F.~Taghavi-Shahri and F.~Arash,
Phys.\ Rev.\ C {\bf 82}, 035205 (2010)
doi:10.1103/PhysRevC.82.035205
[arXiv:1010.1835 [hep-ph]].



\bibitem{Shoeibi:2017lrl} 
S.~Shoeibi, H.~Khanpour, F.~Taghavi-Shahri and K.~Javidan,
Phys.\ Rev.\ D {\bf 95}, no. 7, 074011 (2017)
doi:10.1103/PhysRevD.95.074011
[arXiv:1703.04369 [hep-ph]].



\bibitem{Kovchegov:2017lsr} 
Y.~V.~Kovchegov, D.~Pitonyak and M.~D.~Sievert,
arXiv:1706.04236 [nucl-th].




\bibitem{Aad:2013iua} 
G.~Aad {\it et al.} [ATLAS Collaboration],
Phys.\ Lett.\ B {\bf 725}, 223 (2013)
doi:10.1016/j.physletb.2013.07.049
[arXiv:1305.4192 [hep-ex]].

		
\bibitem{Sadykov:2014aua} 
R.~Sadykov,
arXiv:1401.1133 [hep-ph].
		
		


\bibitem{Giuli:2017tst} 
F.~Giuli,
arXiv:1705.08201 [hep-ph].




\bibitem{Giuli:2017oii} 
F.~Giuli {\it et al.} [xFitter Developers' Team],
Eur.\ Phys.\ J.\ C {\bf 77}, no. 6, 400 (2017)
doi:10.1140/epjc/s10052-017-4931-5
[arXiv:1701.08553 [hep-ph]].



\bibitem{Carrazza:2013bra} 
S.~Carrazza [NNPDF Collaboration],
PoS DIS {\bf 2013}, 279 (2013)
[arXiv:1307.1131 [hep-ph]].



\bibitem{Slominski:2005bw} 
W.~Slominski, H.~Abramowicz and A.~Levy,
Eur.\ Phys.\ J.\ C {\bf 45}, 633 (2006)
doi:10.1140/epjc/s2005-02458-7
[hep-ph/0504003].



\bibitem{Martin:1998sq} 
  A.~D.~Martin, R.~G.~Roberts, W.~J.~Stirling and R.~S.~Thorne,
  Eur.\ Phys.\ J.\ C {\bf 4}, 463 (1998)
  doi:10.1007/s100529800904, 10.1007/s100520050220
  [hep-ph/9803445].
  
  
  
\bibitem{Martin:2004dh} 
  A.~D.~Martin, R.~G.~Roberts, W.~J.~Stirling and R.~S.~Thorne,
  Eur.\ Phys.\ J.\ C {\bf 39}, 155 (2005)
  doi:10.1140/epjc/s2004-02088-7
  [hep-ph/0411040].
  

		
		

\bibitem{Bertone:2013vaa} 
V.~Bertone, S.~Carrazza and J.~Rojo,
Comput.\ Phys.\ Commun.\  {\bf 185}, 1647 (2014)
doi:10.1016/j.cpc.2014.03.007
[arXiv:1310.1394 [hep-ph]].





\bibitem{deFlorian:2015ujt} 
D.~de Florian, G.~F.~R.~Sborlini and G.~Rodrigo,
Eur.\ Phys.\ J.\ C {\bf 76}, no. 5, 282 (2016)
doi:10.1140/epjc/s10052-016-4131-8
[arXiv:1512.00612 [hep-ph]].





\bibitem{Hwa:1979bx} 
R.~C.~Hwa,
Phys.\ Rev.\ D {\bf 22}, 759 (1980).
doi:10.1103/PhysRevD.22.759




\bibitem{Hwa:1994uha} 
R.~C.~Hwa,
Phys.\ Rev.\ D {\bf 51}, 85 (1995).
doi:10.1103/PhysRevD.51.85



\bibitem{Hwa:2002zu} 
R.~C.~Hwa and C.~B.~Yang,
Phys.\ Rev.\ C {\bf 66}, 025205 (2002)
doi:10.1103/PhysRevC.66.025205
[hep-ph/0204289].




\bibitem{Carrazza:2015dea} 
S.~Carrazza,
arXiv:1509.00209 [hep-ph].


\bibitem{Mottaghizadeh:2016krr} 
M.~Mottaghizadeh, P.~Eslami and F.~Taghavi-Shahri,
Int.\ J.\ Mod.\ Phys.\ A {\bf 32}, no. 14, 1750065 (2017)
doi:10.1142/S0217751X17500658
[arXiv:1607.07754 [hep-ph]].




\bibitem{Roth:2004ti} 
M.~Roth and S.~Weinzierl,
Phys.\ Lett.\ B {\bf 590}, 190 (2004)
doi:10.1016/j.physletb.2004.04.009
[hep-ph/0403200].



\bibitem{Mottaghizadeh:2017vef} 
  M.~Mottaghizadeh, F.~T.~Shahri and P.~Eslami,
  arXiv:1707.00108 [hep-ph].




\bibitem{Arash:2007wn} 
F.~Arash and F.~Taghavi-Shahri,
JHEP {\bf 0707}, 071 (2007)
Erratum: [JHEP {\bf 1008}, 106 (2010)]
doi:10.1088/1126-6708/2007/07/071, 10.1007/JHEP08(2010)106
[arXiv:0708.1801 [hep-ph]].




\bibitem{Arash:2006nw} 
F.~Arash,
Nucl.\ Phys.\ Proc.\ Suppl.\  {\bf 152}, 92 (2006).
doi:10.1016/j.nuclphysbps.2005.08.018




\bibitem{Yazdi:2014zaa} 
Z.~Alizadeh Yazdi, F.~Taghavi-Shahri, F.~Arash and M.~E.~Zomorrodian,
Phys.\ Rev.\ C {\bf 89}, no. 5, 055201 (2014)
doi:10.1103/PhysRevC.89.055201
[arXiv:1401.1295 [hep-ph]].



\bibitem{Rinaldi:2016mlk} 
M.~Rinaldi and F.~A.~Ceccopieri,
Phys.\ Rev.\ D {\bf 95}, no. 3, 034040 (2017)
doi:10.1103/PhysRevD.95.034040
[arXiv:1611.04793 [hep-ph]].




\bibitem{Gaunt:2009re} 
J.~R.~Gaunt and W.~J.~Stirling,
JHEP {\bf 1003}, 005 (2010)
doi:10.1007/JHEP03(2010)005
[arXiv:0910.4347 [hep-ph]].


\end{thebibliography}
%


%

\end{document}